\documentclass[prc,preprint,nofootinbib,tightenlines]{revtex4}
\usepackage{graphicx}  %
\usepackage{bm}  %                                          

\newcommand{\beq}{\begin{equation}}
\newcommand{\eeq}{\end{equation}}
\newcommand{\bqa}{\begin{eqnarray}}
\newcommand{\eqa}{\end{eqnarray}}

\def\square{\vcenter{\vbox{\hrule height.4pt
          \hbox{\vrule width.4pt height8pt
          \kern8pt\vrule width.4pt}\hrule height.4pt}}}

\begin{document}

\title{On the Convergence of the Linear $\delta $ Expansion \\
for the Shift in $T_{c}$ for Bose-Einstein Condensation}
\author{Eric Braaten and Eugeniu Radescu}
\affiliation{Physics Department, Ohio State University, Columbus OH 43210, USA}

\begin{abstract}
The leading correction from interactions to the transition temperature $T_{c}$ 
for Bose-Einstein condensation can be obtained from a
nonperturbative calculation in the critical $O(N)$ scalar field theory in 3
dimensions with $N=2$. We show that the linear $\delta $ expansion can
be applied to this problem in such a way that in the large-$N$ limit it
converges to the exact analytic result. 
If the principal of minimal sensitivity is used to optimize the
convergence rate, the errors seem to decrease exponentially
with the order in the $\delta$ expansion.
For $N=2,$ we calculate the shift in 
$T_{c}$ to fourth order in $\delta $. The results are consistent with slow
convergence to the results of recent lattice Monte Carlo calculations.
\end{abstract}

\maketitle

\newpage

\section{Introduction}

The critical temperature $T_{c}$ for Bose-Einstein condensation in an ideal
gas of identical bosons was deduced in 1924 by Albert Einstein 
\cite{Einstein}:
\begin{equation}
k_{B}T_{c}= {2\pi \over \zeta ({3\over2})^{2/3}}
\frac{\hbar ^{2}n^{2/3}}{m},  
\label{Einstein}
\end{equation}
where $n$ is the number density of the bosons. An obvious next question is
the shift $\Delta T_{c}$ in the critical temperature due to the interaction
between the bosons. If the potential between two bosons is short-ranged,
their scattering amplitude at sufficiently low energy is a constant
proportional to the $s$-wave scattering length $a.$ The Bose gas with such
an interaction has been studied extensively since the pioneering work 
\cite{Lee1} of Lee and Yang in 1957 on bosons with a hard-sphere potential.
However the shift in $T_{c}$ from such an interaction remained an unsolved
problem until recently.

A major breakthrough was made by Baym et al.~\cite{Baym1}, who showed that the
problem could be reduced to a nonperturbative calculation in an effective 
$3$-dimensional statistical field theory with $O(2)$ symmetry. 
This demonstrated
conclusively that the leading order shift was linear in $a$:
\begin{equation}
{\Delta T_{c} \over T_{c}} = c \, n^{1/3}a,
\label{c-def}
\end{equation}
where $c$ is a numerical constant. 
Using a self-consistent calculation of the propagator,
they estimated the coefficient to be $c=2.9$ \cite{Baym1}.
Baym, Blaizot and Zinn-Justin calculated
the coefficient analytically in the large-$N$ limit
and obtained $c=2.33$ \cite{Baym2}. 
The first correction in the $1/N$ expansion was calculated by
Arnold and Tomasik and it reduces $c$ to 1.72 \cite{Arnold1}. 
Definitive calculations of the coefficient for the physical case $N=2$
were carried out by Kashurnikov, Prokof'ev, and Svistunov \cite{K-P-S} 
and by Arnold and Moore \cite{Arnold2} by applying Monte Carlo methods 
to a lattice formulation of the effective field theory
with the result $c=1.32\pm 0.02$.
The second order correction to $\Delta T_{c}/T_{c}$ 
proportional to $( an^{1/3})^{2}$ has since been calculated
by Arnold, Moore and Tomasik \cite{Arnold3}. 
Before the recent breakthrough, there had been many previous attempts 
to calculate the leading shift in  $T_{c}$ \cite{Stoof,G-L,H-K,W-I-K}. 
The results for the coefficient $c$ in (\ref{c-def})
ranged from $-1$ to $+4.7$.
The definitive solution of this problem provides a way to assess the
reliability of the various calculational methods.

One particularly simple method that can be applied to this problem is the
{\it linear $\delta$ expansion} \cite{DeltaExp},
also known as {\it optimized perturbation theory} \cite{Stevenson:1981vj}
or {\it variational perturbation theory} \cite{Kleinert}. 
With this method, an arbitrary
parameter $m$ is introduced into the theory and calculations are carried
out using perturbation theory in a formal expansion 
parameter $\delta$ whose value is 1. 
The rate of convergence can be improved by adjusting the 
parameter $m$ at each order in $\delta$ by some criterion 
such as the principal of minimal sensitivity.
It was first applied 
to the problem of calculating $\Delta T_c$
by de Souza Cruz, Pinto and Ramos \cite{Ramos}.
At second, third, and fourth orders in $\delta$, 
the predictions are $c=3.06$, $2.45$, and $1.48$, respectively \cite{Ramos2},
which seem to be converging to the lattice Monte Carlo result.

In this paper, we revisit the application of the linear $\delta$ 
expansion to the calculation of the shift in $T_c$.  
In Section II, we review the reduction of the calculation of 
$\Delta T_c$ to a problem in a 3-dimensional statistical field theory. 
In section III, we introduce the linear $\delta$ expansion 
and review the results of Refs.~\cite{Ramos2}. 
We present two alternative precriptions for applying this method to 
the shift in $T_c$.
In Section IV, we show that in the large-$N$ limit, 
these alternative prescriptions converge to the analytic result
of Ref.~\cite{Baym2} for some range of $m$. 
We study the improvement in the convergence rate that is obtained by using
the principal of minimal sensitivity to choose a value of $m$ 
that depends on the order in $\delta$.
We show that there is a family of solutions to this equation that 
seems to improve the convergence rate from a
power law in the order of the $\delta$ expansion to exponential.
In Section V, we calculate the result for $N=2$ 
at $3^{\rm rd}$ and $4^{\rm th}$ orders in $\delta$. 
In Section VI, we compare our results 
with the previous calculations using the 
linear $\delta$ expansion of Refs.~\cite{Ramos,Ramos2}.
Our results are consistent with slow convergence to the 
Monte Carlo results of Refs.~\cite{K-P-S,Arnold2}.

\section{Effective Theory}

We consider a system of identical bosons of mass $m$ at a temperature $T$
and number density $n$
close to the critical temperature $T_{c}$ for Bose-Einstein condensation.
If the effects of interactions are small,
Einstein's formula (\ref{Einstein}) implies that the interparticle 
spacing $n^{-1/3}$ and the thermal wavelength
$\lambda_T =( 2\pi \hbar ^{2}/mk_{B}T) ^{1/2}$ are comparable.
We assume that they are both large compared to the range $R$ of the
two-body potential.  In this case, the interaction between the bosons
can be characterized entirely by the s-wave scattering length $a$,
which we assume to have magnitude of order $R$.
Thus we require $a \ll n^{-1/3},\lambda_T$. 
This system can be described by a second-quantized Schr\"{o}dinger
equation. The corresponding imaginary-time Lagrangian is
\begin{equation}
{\cal L} = \psi^* {\partial  \over \partial \tau }\psi
- {1 \over 2m} \psi^* \nabla^2 \psi 
- \mu \, \psi^* \psi 
+ {2\pi a \over m} (\psi^* \psi )^{2},
\label{L-it}
\end{equation}
where $\mu $ is the chemical potential 
associated with the particle number density. 
In this paper, we generally use units such that $\hbar =k_{B}=1$. 
Treating the system at finite temperature using the imaginary time
formalism, the field $\psi ({\bf x},t)$ can be decomposed into Fourier
modes $\psi _{n}\left( {\bf x}\right) $ with Matsubara frequencies 
$\omega_{n}=2\pi nT.$  The zero frequency mode 
$\psi _{0}({\bf x})$ is defined by
\begin{equation}
\psi _{0}({\bf x}) = T \int_{0}^{1/T} d\tau \, \psi ({\bf x},t).
\end{equation}
If the temperature is near the critical temperature $T_c$,
the chemical potential $\mu$ is small in magnitude compared to $T$.
For momenta small compared to  $(mT)^{1/2}$,
the nonzero frequency modes have Euclidean energies of order $T$,
while the zero frequency modes have Euclidean energies of order $|\mu|$.
Thus at large distances much greater than $\lambda_T$, 
the nonzero modes decouple from the dynamics, 
leaving an effective theory of
only the zero modes \cite{Baym1}. 
A naive derivation of the effective lagrangian is obtained simply by
substituting $\psi ({\bf x},t) \to \psi _{0}({\bf x})$ in (\ref{L-it}).
The action then reduces to
\begin{equation}
\int_{0}^{1/T}d\tau \int d^{3}x\text{ }{\cal L}[\psi ({\bf x},t)]
\longrightarrow 
\int d^{3}x\text{ }{\cal L}_{{\rm eff}}[\psi_{0}({\bf x})],
\end{equation}
where the effective lagrangian is 
\begin{equation}
{\cal L}_{\rm eff} = 
- {1 \over 2mT} \psi_0^* \nabla^2 \psi_0 
- {\mu \over T} \psi_0^* \psi_0 
+ {2\pi a \over m T} (\psi_0^* \psi_0 )^{2}.
\label{L-0}
\end{equation}
It is convenient to replace the complex field $\psi_0$
by the 2-component real field $\vec \phi
=(\phi _{1},\phi _{2})$ defined by 
\begin{equation}
\psi _{0}({\bf x})= (mT)^{1/2} \left( \phi _{1}({\bf x}) + i \phi _{2}({\bf x}) \right).
\end{equation}
The effective lagrangian density for the dimensionally reduced theory can then
be written as
\begin{equation}
{\cal L}_{{\rm eff}}=
-{1 \over 2} \vec \phi \cdot \nabla^2 \vec \phi
+ {1 \over 2} r {\vec \phi}^{\, 2}
+ {1 \over 24} u \left( {\vec \phi}^{\, 2} \right)^{2},
\label{L-eff}
\end{equation}
where the parameters are
\begin{eqnarray}
r &=&-2m\mu , 
\\
u &=&48\pi amT.
\label{u-a}
\end{eqnarray}
This $O(2)$ field theory can be generalized to an
$O(N)$ field theory by replacing $\vec \phi$ by an $N$-component real field
$\vec \phi=(\phi _{1},...,\phi _{N})$.

This naive derivation of the effective lagrangian is supported by a more
sophisticated analysis within the effective field theory framework
\cite{Arnold3}.
This analysis reveals that there are also higher dimension operators 
in the effective lagrangian (\ref{L-eff}), such as $(\vec \phi^{\, 2})^3$ 
and $\nabla^2 \vec \phi \cdot \nabla^2 \vec \phi$, 
but they are not needed to calculate the shift in $T_c$ 
to leading order in $a n^{1/3}$.
It also reveals that there are corrections to the coefficients
$r$ and $u$ that can be calculated perturbatively as expansions
in $a/\lambda_T$.  Again, they are not needed to calculate  
$\Delta T_c$ to leading order in $an^{1/3}$.
Finally, since the scale $T$ has been removed from the problem,
the effective field theory defined by the effective lagrangian
(\ref{L-eff}) requires an ultraviolet cutoff.
Fortunately, the leading order shift in $T_c$ is dominated by 
the infrared region of momentum space and is insensitive to 
the ultraviolet cutoff.

	From Einstein's formula (\ref{Einstein}), the leading order shift 
$\Delta T_{c} $ in the critical temperature at fixed number density 
is related to the leading order shift $\Delta n_{c}$ in the 
critical density at fixed temperature by the simple relation
\begin{equation}
{\Delta T_{c} \over T_{c}}=-\frac{2}{3}\frac{\Delta n_{c}}{n_{c}}.
\label{delTc-1}
\end{equation}
The number density in the original theory is $n=\left\langle \psi
^{\ast }\psi \right\rangle ,$ where the angular brackets indicate the
expectation value in the state corresponding to a homogeneous gas. 
The number density in the effective theory,
to the accuracy required to calculate the shift in $T_c$ 
to leading order in $an^{1/3}$, is 
\begin{equation}
n=n_{0}+mT \, \langle \vec \phi^{\, 2}\rangle,
\label{n}
\end{equation}
where $n_{0}$ is the short-distance 
contribution to the number density, which depends on the ultraviolet cutoff. 
The angular brackets
indicate the statistical average of the operator $\vec \phi^{\, 2}$. 
It can be expressed as 
\begin{equation}
\langle \vec \phi^{\, 2} \rangle =
2\int_{p}\left[p^{2}+r+\Sigma(p)\right] ^{-1},  
\label{fi2}
\end{equation}
where $\Sigma (p)$ is the self-energy of the $\vec \phi $ field. 
The integral over $\bf p$ is divergent and requires an ultraviolet cutoff.
The number density for an ideal gas is given by the same expression
with $a=0,$ which implies $\Sigma(p)=0$:
\begin{equation}
n|_{a=0}=n_{0}+2mT\int_{p}\left[ p^{2}+r\right] ^{-1}.  \label{na0}
\label{n0}
\end{equation}
Again, the integral over $\bf p$ is divergent 
and requires an ultraviolet cutoff.
To the accuracy required to calculate the shift in $T_c$ 
to leading order in $an^{1/3}$, the constant $n_{0}$ is the same
as in (\ref{n}). The critical point is determined by
the condition that the correlation length associated with the propagator 
in (\ref{fi2}) is infinite:
\begin{equation}
r+\Sigma (0)=0.
\end{equation}
This condition reduces to $r=0$ if $a=0$. 
The change $\Delta n_c$ in the critical density at fixed temperature $T$ 
due to interactions is the difference between (\ref{n}) with 
$r =-\Sigma (0)$ and (\ref{n0}) with $r =0$.
The expression (\ref{delTc-1}) for the fractional shift in $T_c$
can therefore be written as
\begin{equation}
{\Delta T_{c} \over T_{c}}= -{2 \over3} {m T \over n} \Delta,
\label{Delta-def}
\end{equation}
where $\Delta$ is a quantity that can be calculated 
within the effective field theory:
\begin{equation}
\Delta = N \int_{p}\left[ \left[ p^{2}+\Sigma(p)-\Sigma(0)\right]
^{-1}-\left( p^{2}\right) ^{-1}\right] .  
\label{Delta-gen}
\end{equation}
The integral over $\bf p$ is convergent 
and therefore no longer requires an ultraviolet cutoff. 
We have replaced the prefactor 2 in (\ref{Delta-gen}) by $N$ 
to generalize it to the $O(N)$ case.
Comparing (\ref{Delta-def}) with (\ref{c-def}) and using 
(\ref{u-a}) and (\ref{Einstein}), we find that the coefficient $c$ is 
\begin{equation}
c = {8 \pi \over 3} \zeta(\mbox{$3\over2$})^{-4/3}
{\Delta |_{N=2} \over (-u/48\pi^2)} .  
\label{c-Delta}
\end{equation}

The effective theory defined by the lagrangian (\ref{L-eff}) is 
superrenormalizable.  Although the theory has ultraviolet divergences, 
$\Delta $ is dominated by infrared physics and it is insensitive to
the ultraviolet cutoff.  The relevant length scales are therefore 
those set by the parameters $r$ and $u$ of the effective field theory.
The critical point can be reached by tuning the parameter $r$.
At the critical point, there is only one relevant length scale 
and it is set by the parameter $u$.  Since $\Delta $ has dimensions
of length, it must be proportional to $u$ by simple dimensional analysis.
The determination of the coefficient of $u$ requires a nonperturbative
calculation.  Since $u$ is linear in $a$ as $a\to 0$,
this demonstrates that the leading term in the shift in $T_c$ 
is linear in $a$. 

The calculation of the first order shift $\Delta T_{c}$ in the critical
temperature has been reduced to the calculation of the quantity $\Delta $ in
(\ref{Delta-gen}) for $N=2.$ The first calculation of this quantity within a
systematically improvable framework was a calculation to leading order in
the $1/N$ expansion by Baym, Blaizot, and Zinn-Justin \cite{Baym2}: 
\begin{equation}
\Delta = - {Nu \over 96\pi ^{2}}
\qquad \left( {\rm large} \, N \right).
\label{largeN-exact}
\end{equation}
The next-to-leading order (NLO) correction in the $1/N$ expansion 
was calculated by Arnold and Tomasik \cite{Arnold1}. 
The first two terms in the $1/N$ expansion are 
\begin{equation}
\Delta = \left( 1-\frac{0.527}{N}+...\right) 
\left( -\frac{Nu}{96\pi ^{2}} \right).
\label{Delta-1/N}
\end{equation}
For the physical case $N=2$, $\Delta $ has been calculated nonperturbatively
by Kashurnikov, Prokof'ev, and Svistunov \cite{K-P-S} 
and by Arnold and Moore \cite{Arnold2} 
using Monte Carlo simulations of a lattice formulation of the
$O(2)$ field theory. 
The two results are consistent, but Ref.~\cite{Arnold2} 
quotes a smaller error:
\begin{equation}
\Delta = \left( 0.568\pm 0.008\right) \left( - \frac{u}{48\pi ^{2}} \right)
\qquad \left( N=2\right).
\label{Delta-MC}
\end{equation}
If we set $N=2$ in the $1/N$ expansion in (\ref{Delta-1/N}), 
the coefficient of $-u/48\pi ^{2}$ is 
$1$ at leading order and $0.74$ at next-to-leading order. 
These coefficients seem to be approaching 
the lattice Monte Carlo result $0.57$.

\section{Linear $\protect\delta $ expansion}

Any attempt to calculate $\Delta T_{c}$ using ordinary perturbation 
theory in $u$ is doomed to failure.
In the symmetric phase where $r>0$, perturbation theory 
is an expansion in powers of $u/r^{1/2}$.
If we naively let $r \to 0$ to find the critical behavior,
the perturbative corrections are infrared divergent, 
with the degree of divergence becoming more severe at each 
successive order.  It is useful to think of perturbation theory 
as an expansion in powers of  $u \xi$,
where $\xi$ is the correlation length. 
Perturbation theory breaks down in the critical region, 
because $u \xi$ is of order 1 and every order of perturbation theory 
gives a leading order contribution.

The {\it linear $\delta$ expansion} (LDE) is a particularly simple
method for obtaining nonperturbative results using perturbative techniques.
Calculations are carried out as in ordinary perturbation theory by
evaluating a finite number of Feynman diagrams. 
The name ``linear $\delta$ expansion'' comes from the fact that 
it can be defined by a lagrangian whose coefficients are linear
in a formal expansion parameter $\delta$.  
If ${\cal L}$ is the lagrangian for the system of interest,
the lagrangian that generates the LDE has the form
\begin{equation}
{\cal L}_{\delta }= (1-\delta ){\cal L}_{0} + \delta \, {\cal L},
\label{L-delta}
\end{equation}
where ${\cal L}_{0}$ is the lagrangian for an exactly solvable theory,
The lagrangian ${\cal L}_{\delta }$ interpolates 
between ${\cal L}_{0}$ when $\delta =0$ and ${\cal L}$ when $\delta =1$.  
The linear $\delta$ expansion is much easier to apply than the 
{\it nonlinear $\delta$ expansion} in which the lagrangian 
that interpolates between ${\cal L}_{0}$ and ${\cal L}$
involves an operator that is
a nonlinear function of $\delta$ \cite{NLDE}.

The lagrangian ${\cal L}_{0}$ for the exactly solvable theory
typically involves an arbitrary parameter $m$.
At $\delta =1$, the lagrangian (\ref{L-delta}) is independent of $m$.
If a perturbation series in $\delta$ converges at $\delta =1$,
it must converge to a value that is independent of $m$.
However at any finite order in the $\delta$ expansion, 
results depend on $m$.
As $m$ varies over its physical range,
the prediction for the observable often extends out to $\pm \infty$.
Thus some prescription 
for $m$ is required to obtain a definite prediction.
A simple prescription for $m$ is the 
{\it principle of minimal sensitivity} (PMS)
that the derivative with respect to $m$ should vanish.
The LDE supplemented by the PMS prescription is often referred to as 
{\it optimized perturbation theory} \cite{Stevenson:1981vj} 
or {\it variational perturbation theory} \cite{Kleinert}.
A prescription such as PMS may improve the convergence 
rate of the LDE, because it  allows the variable $m$ 
to change with the order $n$ of the LDE.

The convergence properties of the LDE have been studied extensively 
for the quantum mechanics problem of the anharmonic oscillator.
The hamiltonian is
\begin{equation}
H = {1 \over 2} p^2 + {1 \over 2} \omega^2 q^2 + g q^4.
\label{H-delta}
\end{equation}
The LDE is generated by the hamiltonian 
$H_{\delta }= (1-\delta )H_{0} + \delta \, H$, which interpolates between
$H_{0}$ at $\delta=0$ and $H$ at $\delta=1$.  It can be expressed as 
$H = H_0 + H_{\rm int}$, where 
\begin{eqnarray}
H_0 &=& {1 \over 2} p^2 + {1 \over 2} m^2 q^2,
\\
H_{\rm int} &=& -{1 \over 2} \delta (m^2 - \omega^2) q^2 + \delta \, g q^4.
\end{eqnarray}
The LDE has been proven to converge for appropriate order-dependent
choices of the variable $m$ 
that include the PMS criterion as a special case.
Duncan and Jones showed that the finite temperature partition function
converges exponentially, with the errors at $n^{\rm th}$ order
decreasing as $\exp(-b Tn^{2/3}/g^{1/3})$ 
where $b$ is a numerical constant \cite{Duncan-Jones}.
Guida, Konishi, and Suzuki proved that the energy eigenvalues $E_n$ 
converge uniformly in $g$ as $n \to \infty$ \cite{G-K-S}.
Janke and Kleinert showed that the leading coefficient in the 
strong-coupling expansion for the ground state energy $E_0$
converges exponentially, with the errors decreasing like 
$\exp(-c n^{1/3})$ where $c$ is a numerical constant \cite{Janke-Kleinert}.

Bellet, Garcia, and Neveu have made numerical studies of the LDE 
optimized by the PMS criterion for the anharmonic oscillator \cite{B-G-N}.
The PMS criterion is a polynomial equation in the variable $m$, 
so its solutions are in general complex-valued.
The authors of Ref.~\cite{B-G-N} found that the solutions at 
different orders $n$ in the LDE can be organized into distinct families.
The most accurate predictions for the ground state energy $E_0$ 
and for the quantity $\langle q^2 \rangle$ are obtained by using families 
of complex solutions.  The resulting values of $E_0$ 
and $\langle q^2 \rangle$ are of course complex-valued,
but the imaginary parts are small. The authors of Ref.~\cite{B-G-N}
suggested that the real parts could be taken as the predictions 
of the LDE and the imaginary parts could be used as an indication
of the error.

The anharmonic oscillator can be regarded as a Euclidean field theory 
with a single real-valued field in 1 space dimension.
The statistical field theory defined by the lagrangian (\ref{L-eff}) 
is a generalization to a multicomponent field in 3 space dimensions. 
The most serious obstacles to generalizing the convergence proofs 
for the anharmonic oscillator to this more complicated problem
come from the infrared and ultraviolet regions of momentum space.
It is reasonable to expect the convergence behavior to be similar if
appropriate infrared and ultraviolet cutoffs are imposed on the field theory.
The quantity $\Delta$ defined in (\ref{Delta-gen}) is insensitive to 
the ultraviolet region, so we do not expect any complications
from taking the ultraviolet momentum cutoff to $\infty$.
However, $\Delta$ is very sensitive to the infrared region, 
so convergence in the presence of an infrared cutoff gives no information
about the behavior of the LDE in the limit when the infrared 
momentum cutoff goes to 0.

We proceed to apply the LDE to the $O(N)$ statistical field theory
defined by the lagrangian ${\cal L}_{\rm eff}$ in (\ref{L-eff}).   
Our choice for the solvable field theory is the free field theory 
with mass $m$.
The lagrangian ${\cal L}_{\delta }$ defined by 
setting ${\cal L}={\cal L}_{\rm eff}$ in (\ref{L-delta}) can be
written ${\cal L}={\cal L}_0 + {\cal L}_{\rm int}$, where
\begin{eqnarray}
{\cal L}_{0} &=&
-\frac{1}{2}\vec \phi \cdot \nabla^2  \vec \phi
+\frac{1}{2}m^{2} \vec \phi^{\, 2}.
\label{L-zero}
\\
{\cal L}_{\rm int} &=&
{\delta  \over 2} \left( r-m^{2}\right)  \vec \phi^{\, 2}
+{\delta \, u \over 24} \left( \vec \phi^{\, 2} \right)^{2}.
\label{L-int}
\end{eqnarray}
Calculations are carried out by using $\delta$ as a formal expansion 
parameter, expanding to a given order in $\delta$,
and then setting $\delta =1$. 
Thus the Feynman diagrams generated by the LDE will involve 
a propagator with mass parameter $m^2$, 
a 4-point vertex proportional to $\delta \, u$, 
and a mass counterterm proportional to $\delta (r-m^2)$.

In order to apply the LDE to the shift in $T_c$,
we need a prescription for generalizing the quantity $\Delta$ 
defined in (\ref{Delta-gen}) to the field theory whose
lagrangian is ${\cal L}_{\delta }={\cal L}_{0}+{\cal L}_{\rm int}$.
The prescription must have a 
well-defined expansion in powers of $\delta$, and it must reduce to 
(\ref{Delta-gen}) when $\delta = 1$.  One possibility, 
which we will refer to below as prescription II, 
is to use the expression (\ref{Delta-gen}), where $\Sigma(p)$ 
is the self-energy for the field theory defined by 
${\cal L}_\delta$.
Any such prescription defines $\Delta(u,m,\delta)$ as a function of 
three variables.  The $n^{\rm th}$ order approximation in the LDE is 
then obtained by truncating the expansion in powers of $\delta$ at
$n^{\rm th}$ order to obtain a function $\Delta^{(n)}(u,m,\delta)$
and then setting $\delta = 1$:
\begin{equation}
\Delta \approx \Delta^{(n)}(u,m,\delta=1).
\label{Delta-n}
\end{equation}
This sequence of approximations may not converge as $n \to \infty$, 
but if it does converge, we expect it to converge to the correct 
value $\Delta$.

At any finite order $n$, the approximation (\ref{Delta-n}) depends on $m$.  
We will see that as $m$ varies over its physical range from 0 to $+\infty$, 
the range of the predictions for 
$\Delta$ extends out to $+\infty$ for $n$ even and out to $-\infty$
for $n$ odd.  Thus some prescription 
for $m$ is required to obtain a definite prediction.
One prescription is the 
{\it principle of minimal sensitivity} (PMS):
\begin{equation}
{\rm PMS}: \ {d \ \over dm} \Delta^{(n)}(u,m,\delta =1) =0.  
\label{PMS}
\end{equation}
After setting $\delta =1$,
$\Delta^{(n)}$ is a function of $u$ and $m$ only.
By dimensional analysis, a solution $m$ to the PMS criterion
must be proportional to $u$ and the value of $\Delta^{(n)}$
at such a point must also be proportional to $u$. 
An alternative prescription is the {\it fastest apparent convergence} 
(FAC) criterion:
\begin{equation}
{\rm FAC}: \ \Delta^{(n)}(u,m,\delta =1) 
           - \Delta^{(n-1)}(u,m,\delta =1)= 0.  
\label{FAC}
\end{equation}
By allowing the variable $m$ to change with the order $n$, 
prescriptions such as PMS or FAC may improve the convergence 
rate of the LDE.

The LDE was first applied to the problem 
of calculating the shift in $T_c$ 
by de Souza Cruz, Pinto and Ramos \cite{Ramos}. 
Their prescription for 
the quantity $\Delta$ defined in (\ref{Delta-gen}) is
\begin{equation}
\Delta_{\rm I} =N\int_{p}\left[ \left[ p^{2}+m^{2}(1-\delta )+\Sigma(p)-%
\Sigma(0)\right] ^{-1}-\left( p^{2}\right) ^{-1}\right] ,
\label{D-I}
\end{equation}
where $\Sigma(p)$ is the self-energy for the field theory 
whose lagrangian is ${\cal L}_\delta$. 
This prescription agrees with (\ref{Delta-gen}) at $\delta =1$.
Note that in addition to the explicit dependence on $\delta$
in (\ref{D-I}), there is also implicit dependence on $\delta$ 
through $\Sigma(p)$, which can be calculated as a power series in $\delta$.
The subscript I on $\Delta$ in  (\ref{D-I})
distinguishes this prescription 
from alternative prescriptions for $\Delta$ that we will label II and III.
In the Feynman diagrams for evaluating $\Sigma(p)$
in the LDE, all the propagators have mass $m$.
Thus when the prescription I for $\Delta$ is expanded in powers of 
$\delta$, all the integrals involve only propagators of mass $m$,
with the exception of the subtraction term in (\ref{D-I}),
which involves a massless propagator.
The quantity $\Delta$ was calculated to $2^{\rm nd}$ order in $\delta $ 
in Ref.~\cite{Ramos}\ and to $3^{\rm rd}$ and $4^{\rm th}$ order 
in Ref.~\cite{Ramos2}.  After truncating the expansion in powers
of $\delta$ and setting $\delta =1$, they determined $m$ by applying
the PMS prescription (\ref{PMS}).
At $n^{\rm th}$ order in the $\delta $ expansion, this reduces to a polynomial
equation in $m$ with $n$ roots, most of which are complex-valued. 
For a particular choice of the roots at $2^{\rm nd},$ $3^{\rm rd}$,
and $4^{\rm th}$ order in $\delta$, 
the real parts of $\Delta /(-u/48\pi ^{2})$ are $1.31,$
$1.05$, and $0.68,$ respectively. This sequence seems to be approaching
the lattice Monte Carlo result $0.57$. 
In Ref.~\cite{Ramos2}, the authors also applied the LDE
to the second order shift in $T_{c}$ with encouraging results.

Since the result for $\Delta$ is known analytically in the large $N$ limit,
one would also like to compare the predictions of the LDE to this analytic
result.  Unfortunately, as we shall see,  
the prescription I of Ref.~\cite{Ramos2} is not
well-behaved in this limit. We therefore present two alternative
prescriptions for $\Delta$ that are  well-behaved in the large-$N$ limit:
\begin{eqnarray}
\Delta_{\rm II} &=& N\int_{p}\left[ \left(
p^{2}+\Sigma(p)-\Sigma(0)\right) ^{-1}-\left( p^{2}\right)
^{-1}\right] ,  
\label{D-II} 
\\
\Delta_{\rm III} &=& N\int_{p}\left[ \left(
p^{2}+m^{2}\left( 1-\delta \right) +\Sigma(p)-\Sigma(0)\right)
^{-1}-\left( p^{2}+m^{2}\left( 1-\delta \right) \right) ^{-1}\right] .
\label{D-III}
\end{eqnarray}
These both agree with (\ref{Delta-gen}) at $\delta =1$.  
With prescription III, the LDE for $\Delta$ generates 
integrals involving only propagators of mass $m$.  
With prescription II, the integrals involve 
both  propagators of mass $m$ and massless propagators.  

Our prescription III in (\ref{D-III}) is closely related to the 
prescription I in Ref.~\cite{Ramos2}. 
The difference between them is given by a simple integral 
that is evaluated in (\ref{J0}):
\begin{equation}
\Delta_{\rm I} - \Delta_{\rm III}
= -\frac{N}{4\pi }m \sqrt{1-\delta } .
\label{I-III}
\end{equation}
Because of the additional term in $\Delta_{\rm I}$
proportional to $N \sqrt{1-\delta }$, the limit $\delta\to 1$
does not commute with the large-$N$ limit. At $n^{\rm th}$ order in the 
$\delta$ expansion, the terms that dominate as $N\rightarrow \infty$ 
are the $Nm$ term and the highest order term in the $\delta $ expansion 
which is proportional to $N^{n}u^{n}/m^{n-1}$.  The PMS criterion gives a 
value of $m$ that scales like $N^{1-1/n}u$ and the corresponding value of 
$\Delta $ scales like $N^{2-1/n}u.$ 
The correct result is that $\Delta$ scales like $Nu$ in the large $N$ limit.
Thus prescription I has the wrong behavior in the large-$N$ limit 
at any order in the LDE.

\section{Large $\protect N$ limit}

In this section, we study the behavior of the 
linear $\delta$ expansion (LDE) in the large-$N$ limit 
defined by $N\to \infty$, $u \to 0$ with $N u$ fixed.
We calculate $\Delta$ to all orders in  $\delta$ using 
prescriptions II and III.  We show that the LDE converges to the correct
analytic result for $m/Nu$ greater than some minimum value,
with errors that seem to decrease roughly like $n^{-1/3}$.
We also study the improvement in the convergence obtained 
by using the PMS criterion for $m$.
If we use the family of real solutions of the PMS criterion,
the errors in $\Delta$ seem to decrease like $n^{-1/2}$.
If we use a particular family of complex solutions,
the errors seem to decrease exponentially with $n$.

%%%%%%%%%%%%%%%%%%%%%%%%%%%%%%%%%%%%%%%%%%%%%%%%%%%%%%%%%%%%%%%%%%%%%%%%%
\begin{figure}[tbp]
\begin{center}
\centerline{\includegraphics[width=3cm,angle=0,clip=true]{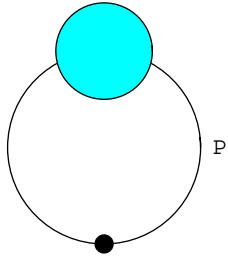}}
\end{center}
\vspace*{-18pt}
\caption{Diagram for $\langle \vec \phi^{\, 2}\rangle$ at the critical point. 
The black blob represents the $\vec \phi^{\, 2}$ vertex, 
while the shaded blob represents the complete propagator 
with subtracted self-energy $\Sigma(p)-\Sigma(0)$.}
\label{fig:Delta}
\end{figure}
%%%%%%%%%%%%%%%%%%%%%%%%%%%%%%%%%%%%%%%%%%%%%%%%%%%%%%%%%%%%%%%%%%%%%%%%%

%%%%%%%%%%%%%%%%%%%%%%%%%%%%%%%%%%%%%%%%%%%%%%%%%%%%%%%%%%%%%%%%%%%%%%%%%
\begin{figure}[tbp]
\begin{center}
\centerline{\includegraphics[width=4cm,angle=0,clip=true]{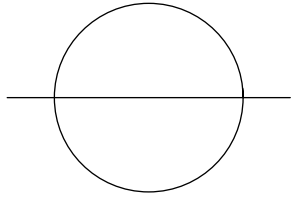}}
\end{center}
\vspace*{-18pt}
\caption{The diagram that contributes to $\Sigma(p)-\Sigma(0)$ 
	at order $\delta^2$.}
\label{fig:2loop}
\end{figure}
%%%%%%%%%%%%%%%%%%%%%%%%%%%%%%%%%%%%%%%%%%%%%%%%%%%%%%%%%%%%%%%%%%%%%%%%%

\begin{figure}[tbp]
\begin{center}
\centerline{\includegraphics[width=5cm,angle=0,clip=true]{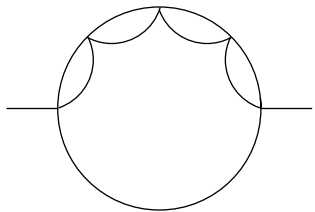}}
\end{center}
\vspace*{-18pt}
\caption{The 4$^{\rm th}$ in the series of diagrams for $\Sigma(p)$
	that survive in the large-$N$ limit.}
\label{fig:largeN}
\end{figure}

The quantity $\Delta$ defined in (\ref{Delta-gen})
has the diagrammatic representation shown in Fig.~\ref{fig:Delta},
where the line with the shaded blob represents the
propagator $[p^{2}+\Sigma(p)-\Sigma(0)]^{-1}$.
In the large-$N$ limit, the leading contribution to 
the subtracted self-energy $\Sigma(p)-\Sigma(0)$ 
comes from the series of bubble-chain diagrams 
whose $1^{\rm st}$ and $4^{\rm th}$ members are shown in 
Figs.~\ref{fig:2loop} and \ref{fig:largeN}. Since these diagrams 
are of order $1/N$, the leading contribution at large $N$
is obtained by expanding the expression (\ref{Delta-gen}) for $\Delta$
to first order in $\Sigma(p)-\Sigma(0)$.
For the prescription II given in (\ref{D-II}), the resulting expression is
\begin{equation}
\Delta_{\rm II} =-N\int_{p}\frac{\Sigma(p)-\Sigma(0)}{\left(
p^{2}\right) ^{2}}.
\label{Delta-largeN}
\end{equation}
The corresponding expression for the prescription III 
is obtained by replacing $p^2 \to p^2+m^2(1-\delta)$ 
in the denominator of (\ref{Delta-largeN}).
Although $\Sigma(p)$ is ultraviolet divergent, the difference 
$\Sigma(p)-\Sigma(0)$ is finite. 
At any finite order in the LDE, 
$\Sigma(p)-\Sigma(0)$ scales as $p^2$ at 
small $p$ and as $\ln p$ for large $p$, so the
integral in (\ref{Delta-largeN}) is convergent.
In addition to the diagrams for $\Sigma(p)$ generated by the interaction term 
$\delta \, u \, \phi^{4}$ in (\ref{L-int}), we must also take into account 
insertions of $\delta \, r = - \delta \, \Sigma(0)$ 
and insertions of $\delta \, m^{2}$. 
The effect of the $\delta \, r$ insertions is to
replace each 1PI self-energy subdiagram in $\Sigma(p)$
by the subdiagram with the zero-momentum limit subtracted.
The effect of the $\delta \, m^{2}$ insertions 
is to replace $m^2$ by $m^2(1-\delta)$.

%%%%%%%%%%%%%%%%%%%%%%%%%%%%%%%%%%%%%%%%%%%%%%%%%%%%%%%%%%%%%%%%%%%%%%%%%
%%%%%%%%%%%%%%%%%%%%%%%%%%%%%%%%%%%%%%%%%%%%%%%%%%%%%%%%%%%%%%%%%%%%%%%%%

We first consider the prescription II for $\Delta$.  
The expression for the large-$N$ diagram for $\Delta$ 
with $n+1$ loops is given in (\ref{A21}). 
The integral $I_{n}$ is a constant multiplied by $m^{-(n-1)}$. 
After taking into account the $\delta \, m^{2}$ insertions,
the sum of the large-$N$ diagrams is 
\begin{equation}
\Delta_{\rm II} =2\text{ }\sum_{n=2}^{\infty }\left( -\delta \frac{%
N}{6}u\right) ^{n}I_{n}\left( 1-\delta \right) ^{-\left( n-1\right) /2}.
\label{23}
\end{equation}
The integrals $I_{n}$ are expressed as 1-dimensional integrals in 
(\ref{B24}). Inserting these expressions into (\ref{23}),
we obtain
\begin{equation}
\Delta_{\rm II} =\delta \frac{N}{24\pi ^{3}}u\text{ }%
\sum_{n=2}^{\infty }\left( -\frac{\delta }{\sqrt{1-\delta }}\frac{1}{\mu }
\right) ^{n-1}\int_{0}^{\infty }dy\frac{y^{2}}{\left( y^{2}+1\right) ^{2}}
\left[ A\left( y\right) \right] ^{n-1},  
\label{D}
\end{equation}
where $\mu $ is the dimensionless mass parameter 
\begin{equation}
\mu =\frac{48\pi m}{N u}  
\label{mu-largeN}
\end{equation}
and $A(y)$ is the function
\begin{equation}
A(y) ={2 \over y} \arctan {y \over 2}.
\label{A-y}
\end{equation}
The prediction for $\Delta$ at $n^{\rm th}$ order in the $\delta $ expansion 
is obtained by expanding the expression (\ref{D}) as a power series 
in $\delta$, truncating after order $\delta^{n}$, 
and then setting $\delta =1$. 

It is easy to show that if the LDE for (\ref{D}) converges, it
converges to the correct analytic result (\ref{largeN-exact}).  After
interchanging the order of the sum and the integral, the sum can be
evaluated and we obtain
\begin{equation}
\Delta_{\rm II} =-\delta \frac{N}{24\pi ^{3}}u\int_{0}^{\infty }dy
\frac{y^{2}}{\left( y^{2}+1\right) ^{2}}\frac{\delta \text{ }A\left(
y\right) }{\sqrt{1-\delta }\text{ }\mu +\delta \text{ }A\left( y\right) }.
\end{equation}
Upon taking the limit $\delta \rightarrow 1,$ all dependence on the
mass parameter $\mu $ disappears:
\begin{equation}
\Delta_{\rm II} =-\delta \frac{N}{24\pi ^{3}}u\int_{0}^{\infty }dy
\frac{y^{2}}{\left( y^{2}+1\right) ^{2}}.
\end{equation}
After evaluating the integral over 
$y$, our final result is (\ref{largeN-exact}).
Note that no prescription for $\mu$ was required to get this result.
Thus, if the $\delta $ expansion converges at some value of $\mu$, 
it should converge to (\ref{largeN-exact}).

%%%%%%%%%%%%%%%%%%%%%%%%%%%%%%%%%%%%%%%%%%%%%%%%%%%%%%%%%%%%%%%%%%%%%%%%%
\begin{figure}[tbp]
\begin{center}
\centerline{\includegraphics[width=12cm,angle=0,clip=true]{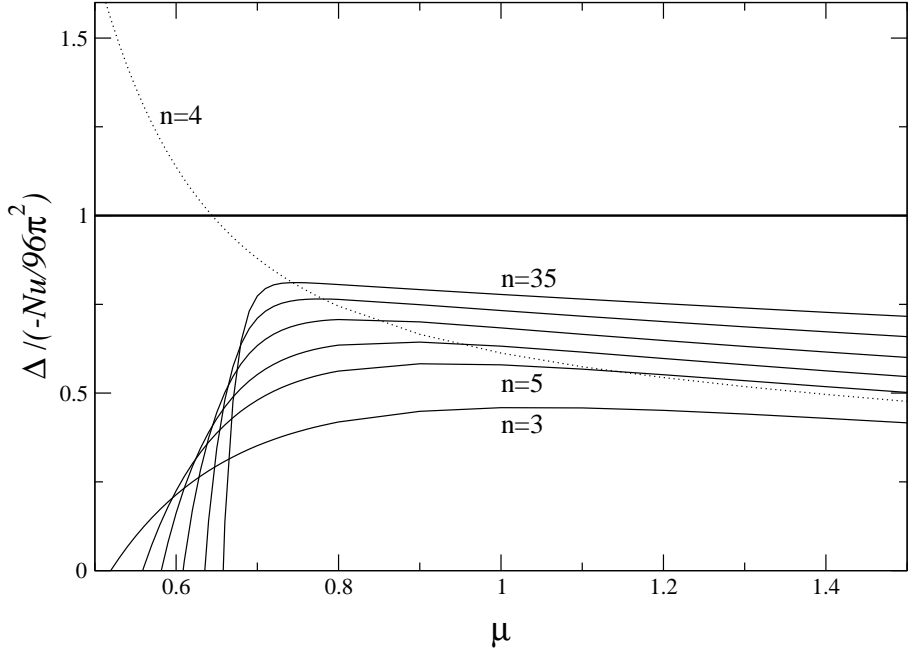}}
\end{center}
\vspace*{-18pt}
\caption{$\Delta/(-Nu/96 \pi^2)$ in the large-$N$ limit as a function of $\mu$ 
	at $n^{\rm th}$ order in the LDE for $n=3$, 4, 5, 7, 11, 19, 35.
	The curves for $n=7$, 11, and 19 
	appear in order between those labelled $n=5$ and 35.}
\label{fig:D-muN}
\end{figure}
%%%%%%%%%%%%%%%%%%%%%%%%%%%%%%%%%%%%%%%%%%%%%%%%%%%%%%%%%%%%%%%%%%%%%%%%%

\begin{figure}[tbp]
\begin{center}
\centerline{\includegraphics[width=12cm,angle=0,clip=true]{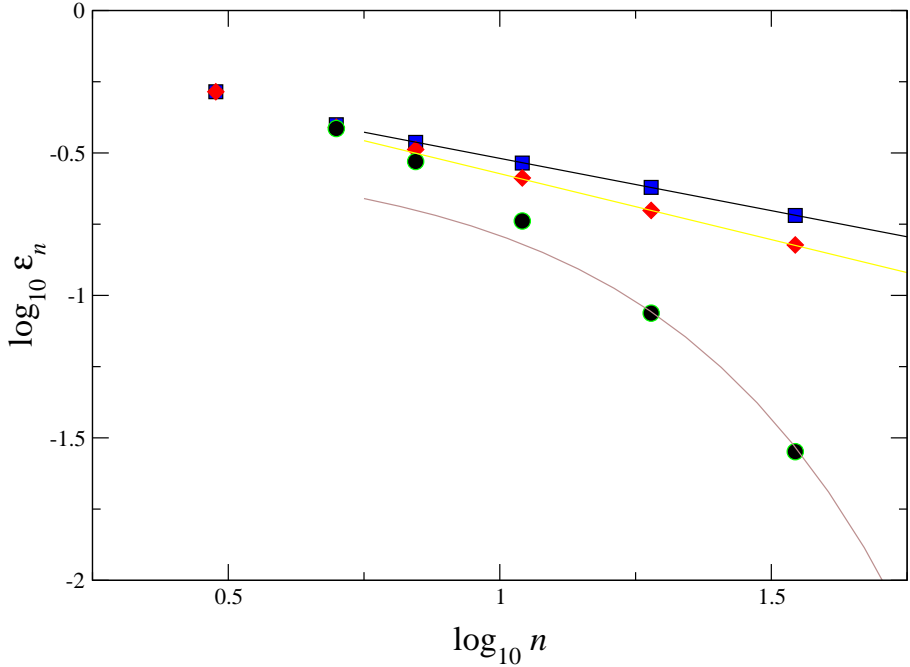}}
\end{center}
\vspace*{-18pt}
\caption{Log-log plot of the fractional error $\varepsilon _n$
	as a function of the order $n$ in the LDE.
	The squares, diamonds, and circles are for $\mu = 1.039$,
	the real solution to the PMS criterion, and the imaginary 
	solution with maximal $|{\rm Im}\Delta|$, respectively.
	The lines are simple curves that pass through the last 2 points.}
\label{fig:err-n}
\end{figure}

The manipulations that were used to show the convergence
to (\ref{largeN-exact}) 
involved several interchanges of limits. 
First, we summed up some terms in the $\delta $
expansion to all orders to obtain the sum over diagrams in (\ref{23}). Next,
we interchanged the order of the sum over diagrams and the integral over the
last momentum for each diagram. 
Next we took the limit $\delta\rightarrow 1$. 
Finally we evaluated the last momentum integral.
It is difficult to translate the conditions for the validity
of each of these steps into a statement about the convergence of the LDE. 
However, the convergence can be easily studied numerically. 
In Fig.~\ref{fig:D-muN}, we show $\Delta$ as a function of the dimensionless
variational parameter $\mu $ for several orders in the $\delta$ expansion: 
$n=3$ and $n=2^{j}+3,$ $j=0,...,5$.
The horizontal line is the analytic result (\ref{largeN-exact}). 
The results are consistent with convergence
to (\ref{largeN-exact}) for all $\mu$ greater than a critical value 
$\mu_c$ which is close to 0.7.  Our best estimate for the endpoint of the
convergence region is $\mu_c = 0.706 \pm 0.003$.
If $\mu < \mu_c$, $\Delta $
seems to diverge to $-\infty $ for $n$ even and $+\infty $ for $n$ odd. 
For any fixed value of $\mu > \mu_c$, 
the convergence with $n$ is very slow. 
In column 1 of Table~\ref{tab1}, we give the values of $\Delta$
at $\mu = 1.039$, which is the location of the minimum of the 
$n=3$ curve in Fig.~\ref{fig:D-muN}. 
We define the fractional error $\varepsilon_n$ by
\begin{equation}
\varepsilon_n = 
{\left|{\rm Re}\Delta - (-Nu/96\pi^2)\right| \over (-Nu/96\pi^2)}.
\label{eps-n}
\end{equation}
(Taking the real part of $\Delta$ is superfluous for real values of $\mu$,
but it will be important later when we consider complex values of $\mu$.)
In Fig.~\ref{fig:err-n}, we show a log-log plot of $\varepsilon_n$
as a function of $n$.  
The points lie close to a straight line, 
indicating that the errors decrease like a power of $n$.
The dotted line is $\varepsilon_n = 0.70 n^{-0.37}$,
which goes through the points for $n=19$ and 35.
Thus the errors seems to decrease roughly like $n^{-1/3}$.

%%%%%%%%%%%%%%%%%%%%%%%%%%%%%%%%%%%%%%%%%%%%%%%%%%%%%%%%%%%%%%%%%%%%%%%%%
%%%%%%%%%%%%%%%%%%%%%%%%%%%%%%%%%%%%%%%%%%%%%%%%%%%%%%%%%%%%%%%%%%%%%%%%%

The rate of convergence can be improved by changing the value of $\mu $ at
each order in the LDE. One possibility is to use the PMS
criterion (\ref{PMS}) to determine $\mu$. At $n^{\rm th}$ order in the $\delta$
expansion, this condition reduces to a polynomial in $\mu $ of order $n-2.$
For $n$ even, there are no real roots. For $n$ odd, there is always one real
root. These roots correspond to the minima of the curves in 
Fig.~\ref{fig:D-muN}.  The values of $\Delta /(-Nu/96\pi^2)$ 
are given in column 2 of Table I. There is slight improvement compared 
to column 1 in the convergence toward the correct value 1.
The fractional errors defined in (\ref{eps-n}) are shown as a function 
of $n$ in Fig.~\ref{fig:err-n}. 
The points lie close to a straight line, 
indicating that the errors decrease like a power of $n$.
The dashed line is $\varepsilon_n = 0.78 n^{-0.46}$,
which goes through the points for $n=19$ and 35.
Thus the errors seem to decrease roughly as $n^{-1/2}$. 

\begin{figure}[tbp]
\begin{center}
\centerline{\includegraphics[width=12cm,angle=0,clip=true]{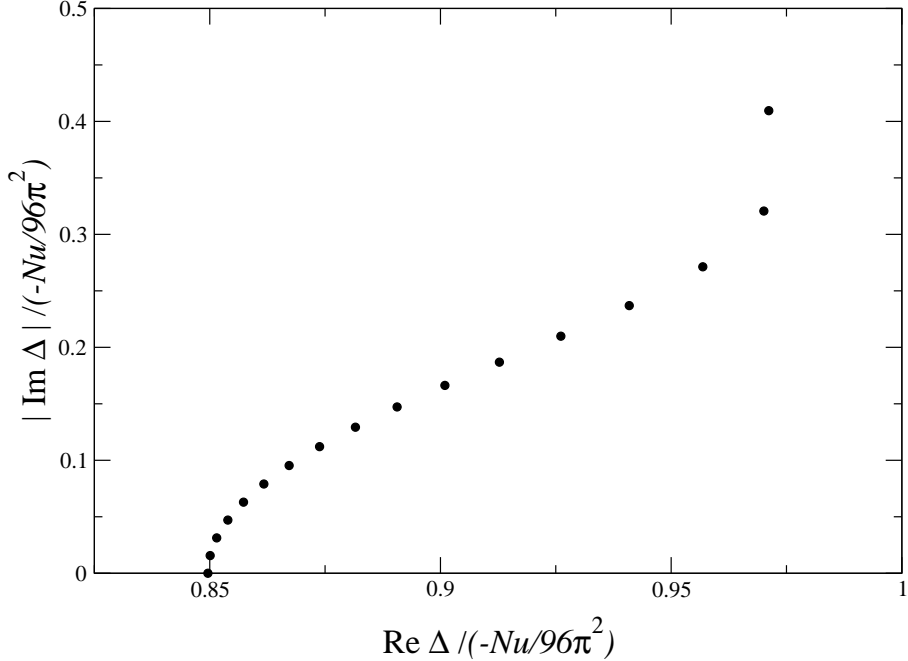}}
\end{center}
\vspace*{-18pt}
\caption{Scatter plot of $|{\rm Im}\Delta|$ vs.~Re$\Delta$ for the solutions 
	$\mu$ of the PMS criterion at $35^{\rm th}$ order in the LDE.}
\label{fig:Im-Re}
\end{figure}

The PMS criterion has either zero or one real solution,
depending on whether $n$ is even or odd.
However there are always $n-2$ complex-valued solutions. 
In the case of the anharmonic oscillator, 
there are families of complex solutions 
with much better convergence properties than families consisting
of purely real solutions \cite{B-G-N}.
In our problem, at any odd order $n,$ the real
solution always has the largest value of Re$\Delta$, which 
is the value farthest from the correct result (\ref{largeN-exact}). 
Thus this family of
solutions gives the slowest possible convergence rate. 
However there is a strong anticorrelation between the errors
in Re$\Delta$ and ${\rm Im} \Delta$.
This is illustrated in Fig.~\ref{fig:Im-Re},
which is a scatter plot of $|{\rm Im}\Delta|$ vs.~Re$\Delta$ 
for the solutions to the PMS criterion for $n=35$. 
The solutions that give
the most accurate values for Re$\Delta$ are those with the largest values for 
$| {\rm Im} \Delta|$.
Thus we can define a nearly optimal family of solutions by choosing those
with the maximal values of $| {\rm Im} \Delta|$.
The values of $\Delta $ for this family of
solutions are given in column 3 of Table I. Note 
that Im$\Delta$ for these solutions shows no sign of converging to $0$,
and in fact iseems to be converging to a value close to 
$0.4 \, {\rm Re} \Delta$.
However the improvement in the convergence of Re$\Delta$ compared 
to the real values of $\Delta$ in column 2 is evident. 
A log-log plot of the fractional errors in
Re$\Delta$ is shown in Fig.~\ref{fig:D-muN}. 
The downward curvature of the points
indicates that the errors decrease faster than any power of $n$.
The solid line is $\varepsilon_n = 0.32(0.93)^n$,
which goes through the points for $n=19$ and 35.
Up to the order to which we have calculated,
the errors are decreasing faster than this exponential.

%%%%%%%%%%%%%%%%%%%%%%%%%%%%%%%%%%%%%%%%%%%%%%%%%%%%%%%%%%%%%%%%%%%%%%%%%
%%%%%%%%%%%%%%%%%%%%%%%%%%%%%%%%%%%%%%%%%%%%%%%%%%%%%%%%%%%%%%%%%%%%%%%%%

We obtain similar results if we use the prescription III for $\Delta$ 
given in (\ref{D-III}). The sum of the large-$N$ diagrams is
the same as in (\ref{23}), except that the integrals $I_n$ are replaced by
$J_n$. The integrals $J_{n}$ are expressed as 1-dimensional integrals in 
(\ref{B25}).  Interchanging the order of the sum and the integral and then
evaluating the sum, we obtain
\begin{equation}
\Delta_{\rm III}=-\delta \frac{N}{16\pi ^{3}}u\int_{0}^{\infty }dy%
\frac{y^{2}}{\left( y^{2}+1\right) \left( y^{2}+4\right) }\frac{\delta \text{
}A\left( y\right) }{\sqrt{1-\delta }\text{ }\mu +\delta \text{ }A\left(
y\right) }.  \label{28}
\end{equation}
Taking the limit $\delta \rightarrow 1,$ all dependence on $\mu$ disappears 
and after evaluating the integral we again obtain (\ref{largeN-exact}).
Thus, if the $\delta $ expansion converges, it should converge to the
correct result. 

The convergence can be again studied numerically. 
We find that it seems to
converge to (\ref{largeN-exact}) for all $\mu > \mu_c$ 
and it seems to diverge for $\mu < \mu_c$, with $\mu_c \approx 0.7$.
In column 4 of Table~\ref{tab1}, we give the values of $\Delta$
for various orders $n$ at $\mu = 0.947$, 
which is the location of the minimum of $\Delta$ for $n=3$.  
The convergence at a fixed value of $\mu$ is very slow.
As in the case of prescription II,
the errors seem to decrease roughly as $n^{-1/3}$.
More rapid convergence can again be obtained by using the PMS criterion
for $\mu $. 
The PMS criterion has no real solutions for $n$
even and 1 real solution for $n$ odd. The values of $\Delta $ for the real
solution at various orders $n$ are given in column 5 of Table I. This is the
family of solutions with the slowest convergence rate for Re$\Delta$.
The errors seem to decrease roughly as $n^{-1/2}$.
A family of solutions with a nearly optimal convergence rate for Re$\Delta$
consists of those with the maximal value of $| {\rm Im} \Delta|$. 
The values of $\Delta $ for this family are shown in column 6 of
Table I. The errors in Re$\Delta$ are numerically larger 
than those with prescription II, but they also seem to decrease 
exponentially with $n$.

\section{Finite $\protect N$}

In this section, we use the linear $\delta$ expansion (LDE) to calculate
$\Delta$ for finite $N$.
We calculate $\Delta$ to $4^{\rm th}$ order in $\delta$ 
using both of the prescriptions II and III.
Setting $N=2$ and using the PMS criterion, we obtain the predictions 
relevant to the shifts in $T_c$ for Bose-Einstein condensation.

%%%%%%%%%%%%%%%%%%%%%%%%%%%%%%%%%%%%%%%%%%%%%%%%%%%%%%%%%%%%%%%%%%%%%%%%%
\begin{figure}[tbp]
\begin{center}
\centerline{\includegraphics[width=12cm,angle=0,clip=true]{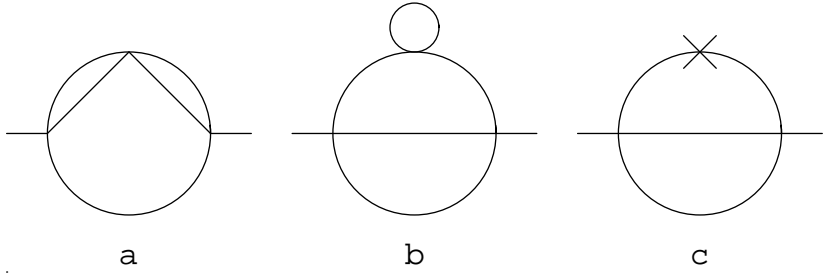}}
\end{center}
\vspace*{-18pt}
\caption{The diagrams that contribute to $\Sigma(p)-\Sigma(0)$ 
	at order $\delta^3$.}
\label{fig:3loop}
\end{figure}
%%%%%%%%%%%%%%%%%%%%%%%%%%%%%%%%%%%%%%%%%%%%%%%%%%%%%%%%%%%%%%%%%%%%%%%%%

\begin{figure}[tbp]
\begin{center}
\centerline{\includegraphics[width=12cm,angle=0,clip=true]{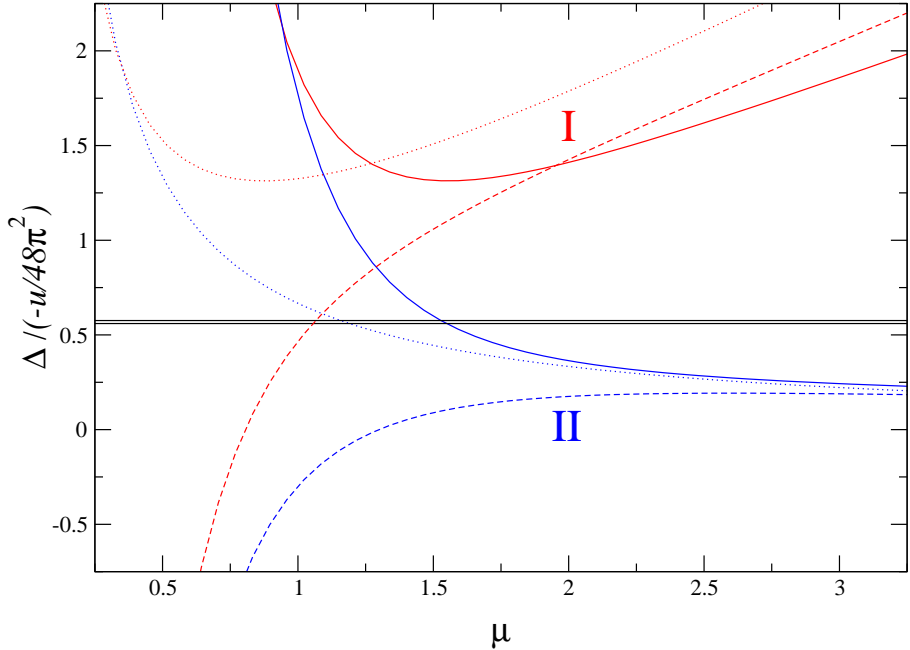}}
\end{center}
\vspace*{-18pt}
\caption{$\Delta/(-u/48 \pi^2)$ for $N=2$ as a function of $\mu$ 
	at $n^{\rm th}$ order in the LDE using prescriptions I
	(upper 3 curves on the right)
	and II (lower 3 curves on the right).  
	The dotted, dashed, and solid curves are for $n=2$, 
	3, and 4, respectively.}
\label{fig:D-mu2}
\end{figure}

The first term in the LDE for $\Delta$ is
second order in $\delta $. It is obtained by calculating the 3-loop
diagram involving the 2-loop self-energy diagram in Fig.~\ref{fig:2loop}. 
The contribution to $\Delta $ of order $\delta ^{3}$ is obtained 
by calculating the 4-loop diagram involving the 
3-loop propagator correction in Fig.~\ref{fig:3loop}. The contribution of
order $\delta ^{4}$ is obtained by calculating the 5-loop diagram
involving the 4-loop propagator corrections in Fig.~\ref{fig:4loop}. 
The diagrams are reduced to
momentum integrals in Appendix \ref{app:diag} 
and the momentum integrals are evaluated in
Appendix \ref{app:int}. The complete expression for $\Delta $ 
for the $O(N)$ theory through order $\delta ^{4}$ is 
\begin{eqnarray}
\Delta_{\rm II} &=&
\delta ^{2}\frac{N(N+2)}{18}u^{2}I_{2}
	\left( 1+\frac{1}{2}\delta +\frac{3}{8}\delta ^{2}\right)
\nonumber \\
&&-\delta^{3}\frac{N(N+2)(N+8)}{108}u^{3}I_{3}
	\left( 1+\delta \right)  
	\nonumber \\
&&+\delta ^{4}\frac{N(N+2)}{648}u^{4} \Bigg[
	(N^{2}+6N+20) I_{4a}  \nonumber 
\\
&&\hspace{3cm}
	+2 (5N+22) \left( I_{4b}+I_{4c}\right) 
	+2(N+2) \left(3I_{4d}+I_{4e}\right) \Bigg].  
\label{Delta-II}
\end{eqnarray}
Inserting the values for the integrals from Appendix B, 
truncating at successively higher orders in $\delta $ 
and then setting $\delta =1$, we obtain the results 
at $2^{\rm nd}$, $3^{\rm rd}$ and $4^{\rm th}$ orders in the LDE:
\begin{eqnarray}
\Delta_{\rm II} ^{(2)}&=&
- \frac{Nu}{96\pi ^{2}}\left( \frac{2}{3} \, \frac{1}{\mu }\right) ,  
\label{d2-II}
\\
\Delta_{\rm II} ^{(3)} &=&
- \frac{Nu}{96\pi ^{2}}\left( \frac{1}{\mu }
- 0.519431\frac{N+8}{N+2}\frac{1}{\mu ^{2}}\right) ,
\label{d3-II} 
\\
\Delta_{\rm II} ^{(4)} &=&
- \frac{Nu}{96\pi ^{2}} \left(
\frac{5}{4}\frac{1}{\mu } - 1.03886\frac{N+8}{N+2}\frac{1}{\mu ^{2}} 
\right.
\nonumber
\\
&&\text{ } \left.
+\left[ 0.430513\frac{N^{2}+6N+20}{\left( N+2\right) ^{2}}
+ 1.09\frac{5N+22}{\left( N+2\right) ^{2}}
- 0.187713\frac{1}{N+2}\right] \frac{1}{\mu ^{3}}
\right),  
\label{d4-II} 
\end{eqnarray}
where $\mu$ is the dimensionless mass parameter 
\begin{equation}
\mu =\frac{48\pi m}{(N+2) u}.  
\label{miu}
\end{equation}
In the large-$N$ limit, this reduces to the mass parameter 
introduced in (\ref{mu-largeN}).
The accuracy of the $4^{\rm th}$ order result in (\ref{d4-II})
is limited by the accuracy to which we have evaluated the integrals 
$I_{4b}$ and $I_{4c}$.
In Fig.~\ref{fig:D-mu2}, we show $\Delta$ for $N=2$ as a function of $\mu$ 
for $n=2,3,4$. The horizontal lines are the edges of the upper and
lower error bars of the lattice Monte Carlo result  
given in (\ref{Delta-MC}).

%%%%%%%%%%%%%%%%%%%%%%%%%%%%%%%%%%%%%%%%%%%%%%%%%%%%%%%%%%%%%%%%%%%%%%%%%
\begin{figure}[tbp]
\begin{center}
\centerline{\includegraphics[width=18cm,angle=0,clip=true]{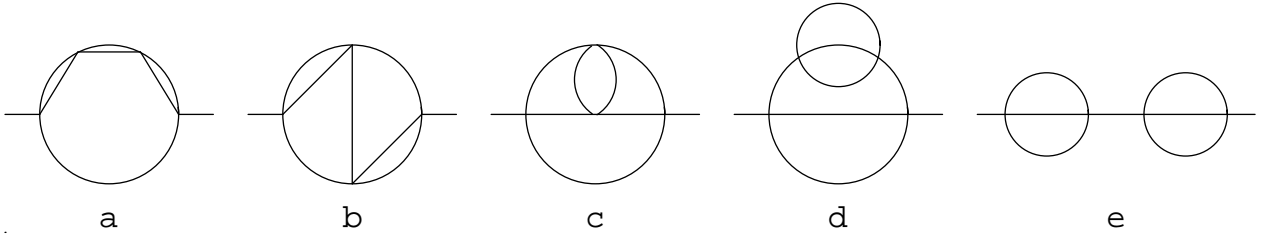}}
\end{center}
\vspace*{-18pt}
\caption{Four-loop diagrams that contribute to $\Sigma(p)-\Sigma(0)$ 
	at order $\delta^4$.}
\label{fig:4loop}
\end{figure}
%%%%%%%%%%%%%%%%%%%%%%%%%%%%%%%%%%%%%%%%%%%%%%%%%%%%%%%%%%%%%%%%%%%%%%%%%
									 
At order $\delta^{2},$ the PMS criterion has no solutions,
because (\ref{d2-II}) is a monotonic function of 
$\mu$.  At order $\delta ^{3},$ the PMS criterion has a single real solution 
$\mu^{(3)}=1.03886(N+8)/(N+2) $ and the
resulting value for $\Delta $ is 
\begin{equation}
\Delta_{\rm II}^{(3)}= 0.481296 \, {N+2 \over N+8} 
\left( -\frac{Nu}{96\pi ^{2}} \right).
\label{Delta3}
\end{equation}
At 4$^{\rm th}$ order in $\delta $, $\Delta $ is a monotonic function for
real-valued $\mu $, so the PMS criterion has no real solutions for $\mu $.
It does however have a conjugate pair of complex solutions 
whose values at $N=2$ are $\mu=2.0772\pm 1.773i$. 
The corresponding values of $\Delta $ are complex,
but we can take the real part as the prediction 
for $\Delta$ at 4$^{\rm th}$ order in the LDE:
\begin{eqnarray}
{\rm Re} \Delta_{\rm II}^{(4)} &=& 0.214 \left( - \frac{u}{48\pi ^{2}} \right)
\qquad \left( N=2\right) .
\label{Delta4:2}
\end{eqnarray}
For $N=2$, the $3^{\rm rd}$ and $4^{\rm th}$ order
approximations in (\ref{Delta3})
and (\ref{Delta4:2}) differ from the lattice Monte Carlo result 
of (\ref{Delta-MC}) by about 66\% and 62\%, respectively.
These predictions are not as accurate as in the large $N$ limit, 
where the errors in the $3^{\rm rd}$ and $4^{\rm th}$ order
approximations are about 52\% and 44\%, respectively.
Furthermore, the percentage decrease in the error
in going from $3^{\rm rd}$ order to $4^{\rm th}$ order
is only about half as large for $N=2$ as it is for large $N$.

%%%%%%%%%%%%%%%%%%%%%%%%%%%%%%%%%%%%%%%%%%%%%%%%%%%%%%%%%%%%%%%%%%%%%%%%%
%%%%%%%%%%%%%%%%%%%%%%%%%%%%%%%%%%%%%%%%%%%%%%%%%%%%%%%%%%%%%%%%%%%%%%%%%

With the prescription III, the expression for $\Delta$
to $4^{\rm th}$ order in $\delta$ is the same as (\ref{Delta-II}),
except that the integrals $I_n$ are replaced by $J_n$.
Inserting the value for the integrals $J_n$ from Appendix B, 
the results at $2^{\rm nd}$, $3^{\rm rd}$ and $4^{\rm th}$ orders 
in the LDE are
\begin{eqnarray}
\Delta_{\rm III} ^{(2)}&=&
- \frac{Nu}{96\pi ^{2}}\left( 0.575364 \, \frac{1}{\mu }\right) ,  
\label{d2-III}
\\
\Delta_{\rm III} ^{(3)} &=&
- \frac{Nu}{96\pi ^{2}}\left( 0.863046 \frac{1}{\mu }
-0.408802 \, \frac{N+8}{N+2}\frac{1}{\mu ^{2}}\right) ,
\label{d3-III} 
\\
\Delta_{\rm III} ^{(4)} &=&
- \frac{Nu}{96\pi ^{2}} \left(
1.07881 \, \frac{1}{\mu } - 0.817603 \frac{N+8}{N+2}\frac{1}{\mu ^{2}} 
\right.
\nonumber
\\
&&\text{ } \left.
+\left[ 0.316944 \, \frac{N^{2}+6N+20}{(N+2)^{2}}
+0.81\frac{5N+22}{(N+2)^{2}}
-0.149896\frac{1}{N+2}\right] \frac{1}{\mu ^{3}}
\right) ,  
\label{d4-III} 
\end{eqnarray}
The accuracy of the $4^{\rm th}$ order result in (\ref{d4-III})
is limited by the accuracy to which we have evaluated the integrals 
$J_{4b}$ and $J_{4c}$.
At order $\delta ^{2},$ the PMS criterion has no solution.  
At order $\delta ^{3},$ it has one real solution for $\mu $.
At 4$^{\rm th}$ order in $\delta $, 
it has a conjugate pair of complex solutions.
The corresponding values of $\Delta$ are given in Table~\ref{tab2}.
The values of $\Delta$ at $3^{\rm rd}$ order and  
of Re $\Delta$ at $4^{\rm th}$ order are close to the corresponding values  
from prescription II, but a little farther from the
lattice Monte Carlo result.

The results of the LDE for prescription I can be obtained from those for
prescription III by using (\ref{I-III}):
\begin{equation}
\Delta_{\rm I}^{(n)}= \Delta_{\rm III}^{(n)} +
\left( -\frac{Nu}{96\pi ^{2}} \right){N+2 \over 2} \, \mu 
\sum_{i=0}^n (-1)^i{{1\over2} \choose i} .
\label{DeltaI-III}
\end{equation}
In Fig.~\ref{fig:D-mu2}, we show $\Delta^{(n)}$ as a function of $\mu$
for $n=2$, 3, and 4.  The values of $\Delta^{(n)}$ at the solutions to the 
PMS criterion are given in column 1 of Table~\ref{tab2}.
At orders 3 and 4, the PMS criterion has a conjugate pair of 
complex solutions.  As the predictions of the LDE,
the authors of Ref.~\cite{Ramos2} chose the positive value for
$\Delta^{(2)}$ and the real parts of the complex values for
$\Delta^{(3)}$ and $\Delta^{(4)}$.  These predictions differ from
the lattice Monte Carlo results by about 131\%, 85\%, and 20\%, respectively.

\section{Conclusions}

The calculation of the shift in $T_c$ to leading order in $a n^{1/3}$ 
can be reduced to a nonperturbative calculation of a quantity
$\Delta$ in a 3-dimensional $O(2)$ field theory at its critical point.
This quantity can be generalized to an $O(N)$ field theory
and it can be calculated systematically using the $1/N$ expansion.
We have shown that the linear $\delta$ expansion (LDE) can be applied to
the calculation of $\Delta$
in such a way that the large $N$ and $\delta \to 1$ limits commute.
The prescriptions II and III defined in (\ref{D-II}) and (\ref{D-III})
have this property, but the prescription I introduced in 
Ref.~\cite{Ramos} does not.

For prescription II and III, we calculated 
$\Delta$ in the large-$N$ limit to all orders in the LDE, 
and studied its convergence properties.
We found that it seems to converge to the correct analytic result
for $\mu > \mu_c$, where $\mu_c \approx 0.7$.  
The convergence for fixed $\mu$ is extremely slow:  
the errors seem to decrease roughly like $n^{-1/3}$.
The convergence can be accelerated by using the PMS criterion
to choose a value of $\mu$ that depends on the order in $\delta$.
The PMS criterion is a polynomial equation with multiple roots, 
so there are various families of predictions for $\Delta$.
The family consisting of the real solutions of the PMS criterion
gives the slowest possible convergence:
the errors in $\Delta$ seem to decrease roughly like $n^{-1/2}$.
The family consisting of the complex solutions of the PMS criterion
with the maximal value of $|{\rm Im}\Delta|$
have nearly optimal convergence:
the errors in Re$\Delta$ seem to decrease exponentially with $n$.
Unfortunately, accurate predictions still require calculations 
to high orders in the $\delta$ expansion.  To achieve 10\%
accuracy requires computing to about $18^{\rm th}$ order
in $\delta$.

For the case $N=2$ relevant to Bose-Einstein condensation,
we calculated $\Delta$ to $4^{\rm th}$ order in the LDE
using both prescriptions II and III.
The quantity $\Delta$ was previously  calculated to $4^{\rm th}$ order 
using prescription I in Ref.~\cite{Ramos2}.  
The PMS criterion has complex solutions beginning at $3^{\rm rd}$ order
for prescription I and at $4^{\rm th}$ order for prescriptions II and III.
We take the values of Re$\Delta^{(n)}$ for these solutions 
to be the predictions for $\Delta$.
The $3^{\rm rd}$ and 4$^{\rm th}$ order predictions from prescription II 
are smaller than the lattice Monte Carlo result by about 66\% and 62\%,
respectively.
The $2^{\rm nd}$, $3^{\rm rd}$ and 4$^{\rm th}$ order predictions 
from prescription I are larger than the lattice Monte Carlo result 
by about 131\%, 85\%, and 20\%, respectively.
We argue that the 20\% error in the 4$^{\rm th}$ order prediction
from prescription I is fortuitously small.  Prescription I 
differs from prescription III by the additive term given in (\ref{I-III}).
The expansion of the factor $\sqrt{1-\delta}$ 
truncated at $n^{\rm th}$ order in $\delta$  and evaluated at $\delta = 1$ 
converges to 0 as $n \to \infty$.  The prediction
from prescription I should therefore approach the prediction 
from prescription III as $n$ increases, which suggests that it will 
converge to the correct answer 
from below.  Because  the $n=2$ prediction
from prescription I is greater than the correct result by about a
factor of 2.3, there must be a
crossover in $n$ where the prediction goes from values greater than the 
correct result to smaller values.  The high accuracy of the 4$^{\rm th}$ 
order result suggests that this crossover occurs at $n=5$.  Thus we expect 
higher order predictions from prescription I to first overshoot 
the correct value and then approach it from below.

Before the lattice Monte Carlo calculation of Ref.~\cite{K-P-S,Arnold2},
there were many previous attempts to calculate the coefficient $c$
in the expression (\ref{c-def}) for the leading order shift 
in $T_c$ for Bose-Einstein condensation.
For most of these methods \cite{Stoof,G-L,H-K,Baym1,W-I-K},
there is no known systematic way of improving the 
calculation to make it more accurate.  
Such a method is best regarded as just a model 
of the critical $O(N)$ field theory whose accuracy can only be determined 
by comparison with a more reliable calculational method. 
A systematically improvable method has two great advantages 
over such a model.  First, the accuracy of the prediction
can be improved with additional effort.
Second, one can make a reliable estimate of the error in the prediction.
One example of a systematically improvable method is the $1/N$ expansion 
\cite{Baym1,Arnold1}, which has been used to 
calculate $\Delta T_c$ to leading order \cite{Baym1} 
and to next-to-leading order \cite{Arnold1}.
Comparison with the lattice Monte Carlo results show that 
these first two approximations
have errors of 77\% and 30\%, respectively.  
Our demonstration that the LDE converges in the large $N$ limit
suggests that the LDE may also be a systematically improvable 
method for $N=2$.

Although the convergence rate of the optimized LDE appears to be 
exponential in the large-$N$ limit, it is still rather slow. 
One must calculate to about $18^{\rm th}$ order in $\delta$ 
to achieve 10\% accuracy.  For the case $N=2$ relevant to
Bose-Einstein condensation, we have calculated $\Delta$ 
to $4^{\rm th}$ order in the LDE.  It may be feasible to extend
the calculation to $5^{\rm th}$ order,
but it would be very difficult to extend it to much higher order.  
The slow convergence in the large-$N$ limit
suggests that even if the LDE also converges for $N=2$, 
a strict expansion in $\delta$ is not useful for quantitative 
calculations.  It may however be possible to use order-dependent 
mappings to change the expansion in $\delta$ into 
a more rapidly converging expansion \cite{Seznec:ev}.

In conclusion, we have shown that the LDE for the quantity $\Delta$ 
that determines the leading order shift in $T_c$ converges 
in the large $N$ limit if we use an appropriate prescription.
If the PMS criterion is used to optimize the convergence,
the errors seem to decrease exponentially in the order of the LDE.
This suggests that the LDE may also be a systematically improvable 
approximation scheme in the case $N=2$ relevant to Bose-Einstein condensation.
With the use of order-dependent mappings, it may be possible to
to develop the LDE into a general and powerful tool for 
quantitative calculations in superrenormalizable field
theories, even at a critical point.

\begin{acknowledgments}
This research was supported by DOE Grant No.~DE-FG02-91-ER4069.
\end{acknowledgments}

%\newpage 

\appendix\renewcommand{\theequation}{\thesection.\arabic{equation}}

\section{Diagrams}
\label{app:diag}

In this appendix, we give the expressions for the diagrams that contribute
to $\Delta $ through order $\delta ^{4}$. The diagrams for $\Delta $
are obtained by inserting propagator corrections into the diagram 
in Fig.~\ref{fig:Delta}, with the propagator corrections subtracted 
at $0$ momentum.  In addition, any 2-loop self-energy subdiagram
like that in Fig.~\ref{fig:2loop} must also be subtracted at $0$ momentum. 

We first introduce a compact notation for the massive propagator and for the
functions of the momentum that correspond to subdiagrams in which 
two points are connected by 2 or 3 lines:
\begin{eqnarray}
D_{1}(p) &=&\frac{1}{p^{2}+m^{2}}, \\
D_{2}(p) &=&\int_{q}\frac{1}{q^{2}+m^{2}}\frac{1}{\left( {\bf p+q}\right)
^{2}+m^{2}}, \\
D_{3}(p) &=&\int_{qr}\frac{1}{q^{2}+m^{2}}\frac{1}{r^{2}+m^{2}}\frac{1}{%
\left( {\bf p+q+r}\right) ^{2}+m^{2}},
\end{eqnarray}
where $\int_{q}$ denotes the 3-dimensional integral over the momentum 
${\bf q:}$
\begin{equation}
\int_{q}=\int \frac{d^{3}q}{(2\pi )^{3}}.
\end{equation}
There is a simple analytic expression for the function $D_{2}(p)$:
\begin{equation}
D_{2}(p)=\frac{1}{4\pi p}\arctan \frac{p}{2m}.
\end{equation}
The function $D_{3}(p)$ is ultraviolet divergent, but the
divergence can be eliminated by a subtraction at $0$ momentum. The
subtracted function can be evaluated analytically:
\begin{equation}
D_{3}(p)-D_{3}\left( 0\right) =\frac{1}{\left( 4\pi \right) ^{2}}\left[ 1-%
\frac{3m}{p}\arctan \frac{p}{3m}-\frac{1}{2}\log \frac{p^{2}+9m^{2}}{9m^{2}}%
\right] .
\end{equation}

We proceed to give the expression for the contributions to $\Delta$
using the prescription II given in (\ref{D-II}). 
The contribution from the 2-loop propagator correction 
in Fig.~\ref{fig:2loop} is
\begin{eqnarray}
\Delta _{2} &=&\frac{1}{6}\delta ^{2}u^{2}\frac{N\left( N+2\right) }{3}I_{2},
\label{A7} \\
I_{2} &=&\int_{p}\frac{1}{\left( p^{2}\right) ^{2}}\left[ D_{3}\left(
p\right) -D_{3}\left( 0\right) \right] .  \label{A8}
\end{eqnarray}
The contribution from the sum of the 3-loop propagator corrections 
in Fig.~\ref{fig:3loop} is
\begin{eqnarray}
\Delta _{3} &=&-\frac{1}{4}\delta ^{3}u^{3}\frac{N\left( N+2\right) \left(
N+8\right) }{27}I_{3},  \label{A9} \\
I_{3} &=&\int_{p}\frac{1}{\left( p^{2}\right) ^{2}}\int_{q}\left[
D_{1}\left( p+q\right) -D_{1}\left( q\right) \right] D_{2}^{2}\left(
q\right) .  \label{A10}
\end{eqnarray}
The contributions from the 4-loop propagator corrections in 
Fig.~\ref{fig:4loop} are
\begin{eqnarray}
\Delta _{4a} &=&\frac{1}{8}\delta ^{4}u^{4}\frac{N\left( N+2\right) \left(
N^{2}+6N+20\right) }{81}I_{4a},  \label{A11} \\
\Delta _{4b} &=&\frac{1}{4}\delta ^{4}u^{4}\frac{N\left( N+2\right) \left(
5N+22\right) }{81}I_{4b}, \\
\Delta _{4c} &=&\frac{1}{4}\delta ^{4}u^{4}\frac{N\left( N+2\right) \left(
5N+22\right) }{81}I_{4c}, \\
\Delta _{4d} &=&\frac{1}{12}\delta ^{4}u^{4}\frac{N\left( N+2\right) ^{2}}{9}
I_{4d}, \\
\Delta _{4e} &=&\frac{1}{36}\delta ^{4}u^{4}\frac{N\left( N+2\right) ^{2}}{9}
I_{4e},
\end{eqnarray}
where the momentum integrals are
\begin{eqnarray}
I_{4a} &=&\int_{p}\frac{1}{\left( p^{2}\right) ^{2}}\int_{q}\left[
D_{1}\left( p+q\right) -D_{1}\left( q\right) \right] D_{2}^{3}\left(
q\right) ,  
\label{A16} 
\\
I_{4b} &=&\int_{p}\frac{1}{\left( p^{2}\right) ^{2}}\int_{qr}D_{1}\left(
q\right) D_{1}\left( q-r\right) D_{2}\left( r\right) 
\nonumber
\\
&& \hspace{3cm} \times 
\left[ D_{1}\left(
r-p\right) D_{2}\left( p-q\right) -D_{1}\left( r\right) D_{2}\left( q\right) 
\right] ,  
\label{A17} 
\\
I_{4c} &=&\int_{p}\frac{1}{\left( p^{2}\right) ^{2}}\int_{qrs}\left[
D_{1}\left( p+q\right) -D_{1}\left( q\right) \right] 
D_{1}\left( r\right) D_{1}\left( q+r\right)
\nonumber
\\
&& \hspace{3cm} \times 
D_{1}\left( s\right) D_{1}\left( q+s\right)
D_{2}\left( r-s\right) ,  
\label{A18} 
\\
I_{4d} &=&\int_{p}\frac{1}{\left( p^{2}\right) ^{2}}\int_{q}\left[
D_{2}\left( p+q\right) -D_{2}\left( q\right) \right] D_{1}^{2}\left(
q\right) \left[ D_{3}\left( q\right) -D_{3}\left( 0\right) \right] ,
\label{A19} 
\\
I_{4e} &=&\int_{p}\frac{1}{\left( p^{2}\right) ^{3}}\left[ D_{3}\left(
p\right) -D_{3}\left( 0\right) \right] ^{2}.  
\label{A20}
\end{eqnarray}

If one uses the prescription III for $\Delta$ given in (\ref{D-II}),
the corresponding
expressions for the diagrams are obtained by replacing each factor of 
$1/p^{2}$ in the integrand by $D_{1}(p)$.
We denote the integral corresponding to $I_n$ by $J_n$. 
If one uses the prescription I of Ref.~\cite{Ramos2},
there is also a diagram that is $0^{\rm th}$ order in $\delta$:
\begin{eqnarray}
\Delta _{0} &=&NJ_{0}, 
\\
J_{0} &=&
\int_{p} \left[ D_{1}(p) - \left( p^{2}\right) ^{-1} \right] .
\label{J0}
\end{eqnarray}

The $4^{\rm th}$ in the class of diagrams that survive in the large-$N$ 
limit is shown in Fig.~\ref{fig:largeN}. 
The expression for the first three diagrams in the series are given
in (\ref{A7}), (\ref{A9}) and (\ref{A11}). The general expression for the
$(n+1)$-loop diagram is
\begin{eqnarray}
\Delta_{n} &=&2C_{n}(N) \left(
-\frac{\delta \text{ }u}{2}\right) ^{n}I_{n},  
\label{A21} 
\\
I_{n} &=&\int_{p}\frac{1}{\left( p^{2}\right) ^{2}}\int_{q}\left[
D_{2}\left( q\right) \right] ^{n-1}\left[ D_{1}\left( p+q\right)
-D_{1}\left( q\right) \right] .  
\label{A22}
\end{eqnarray}
If we use prescription III, the corresponding integral is
\begin{equation}
J_{n} = \int_{p} D_{1}^{2}(p) 
\int_{q} \left[ D_{2}(q) \right]^{n-1}
\left[ D_{1}(p+q) -D_{1}(q) \right] .
\label{A23}
\end{equation}
A closed-form expression for the $O(N) $ factor
associated with this class of diagrams is
\begin{equation}
C_{n}(N) =\left( \frac{N+2}{3}\right) ^{n}+\left( \frac{2}{3}%
\right) ^{n}\frac{\left( N-1\right) \left( N+2\right) }{2},
\end{equation}
where we have chosen the overall normalization such that $%
C_{n}\left( 1\right) =1.$ For $n=2,3$ and $4,$ this expression reproduces
the $O(N) $ factor in (\ref{A7}), (\ref{A9}) and (\ref{A11}).
Note that for $n\geq 3,$ the large-$N$ limit of the coefficient is 
\begin{equation}
C_{n}(N) =\left( \frac{N+2}{3}\right) ^{n}
\left[ 1+{\cal O}(1/N) \right], \qquad n\geq 3.
\end{equation}
The corrections are suppressed by powers of $1/N.$ If we use the simpler
expression $\left( N/3\right) ^{n},$ the error is of order $1/N$ for $%
n\ll N$ and of order $N^{0}$ for $n$ of order $N$.

\section{Integrals}
\label{app:int}

In this appendix, we evaluate the momentum integrals that appear in the
expressions for the diagrams in Appendix A. 

We first consider the integrals $I_n$ that arise if we use prescription II 
for $\Delta$ given in (\ref{D-II}). The 3-loop and 4-loop integrals 
(\ref{A8}) and (\ref{A10}) can be evaluated analytically:
\begin{eqnarray}
I_{2} &=&-\frac{1}{6\left( 4\pi \right) ^{3}}\frac{1}{m}, 
\\
I_{3} &=&-\frac{1}{24\left( 4\pi \right) ^{4}}
\left[ \pi ^{2}+16\log \frac{3}{4}+12Li_{2}\left( -1/3\right) \right] 
\frac{1}{m^{2}}.
\end{eqnarray}
The remaining integrals must be evaluated numerically.
Three of the 5-loop integrals can be reduced to integrals over a single
momentum. The integral $I_{4e}$
in (\ref{A20}) is already in such a form. The integrals $I_{4a}$ in (\ref%
{A16}) and $I_{4d}$ in (\ref{A19}) \ can be reduced to this form by first
integrating over $p:$
\begin{eqnarray}
\int_{p}\frac{1}{\left( p^{2}\right) ^{2}}\left[ D_{1}\left( p+q\right)
-D_{1}\left( q\right) \right] &=&-\frac{m}{4\pi }\frac{1}{\left(
q^{2}+m^{2}\right) ^{2}}, \\
\int_{p}\frac{1}{\left( p^{2}\right) ^{2}}\left[ D_{2}\left( p+q\right)
-D_{2}\left( q\right) \right] &=&-\frac{1}{32\pi ^{2}}\frac{1}{q^{2}+4m^{2}}.
\end{eqnarray}
The resulting expression for the integrals are 
\begin{eqnarray}
I_{4a} &=&-\frac{m}{4\pi }\int_{q}D_{1}^{2}\left( q\right) D_{2}^{3}\left(
q\right) , \\
I_{4d} &=&-\frac{1}{32\pi ^{2}}\int_{q}D_{1}^{2}\left( q\right) \left[
D_{3}\left( q\right) -D_{3}\left( 0\right) \right] \frac{1}{q^{2}+4m^{2}}.
\end{eqnarray}
Reducing these integrals to 1-dimensional integrals and evaluating
then numerically, we obtain
\begin{eqnarray}
I_{4a} &=&\frac{1}{\left( 4\pi \right) ^{6}}\left( -0.338124209\right) \frac{%
1}{m^{3}}, \\
I_{4d} &=&\frac{1}{\left( 4\pi \right) ^{6}}\left( 0.0196337530\right) \frac{%
1}{m^{3}}, \\
I_{4e} &=&\frac{1}{\left( 4\pi \right) ^{6}}\left( 0.0148131532\right) \frac{%
1}{m^{3}}.
\end{eqnarray}
The remaining two 5-loop integrals can be expressed as 3-loop
integrals with the Mercedes-Benz topology. The integral $I_{4b}$ in 
(\ref{A17})\ already has this form. The integral $I_{4c}$ in (\ref{A18})
 can be put in this form by integrating over $p$:
\begin{equation}
I_{4c}=-\frac{m}{4\pi }\int_{qrs}D_{1}^{2}\left( q\right) D_{1}\left(
r\right) D_{1}\left( s\right) D_{1}\left( s-q\right) D_{1}\left( q-r\right)
D_{2}\left( r-s\right) .
\end{equation}
These integrals can be reduced to 7-dimensional integrals that are then
evaluated numerically using the Monte Carlo method to obtain
\begin{eqnarray}
I_{4b} &=&\frac{1}{\left( 4\pi \right) ^{6}}\left( -0.25\right) \frac{1}{%
m^{3}}, \\
I_{4c} &=&\frac{1}{\left( 4\pi \right) ^{6}}\left( -0.18\right) \frac{1}{%
m^{3}}.
\end{eqnarray}

The integrals $J_n$ required if $\Delta$ is calculated 
using the prescription III differ from the integrals $I_{n}$ 
calculated above only in the substitution $1/p^{2}\rightarrow D_{1}(p)$. 
The 3-loop integral can be evaluated analytically:
\begin{equation}
J_{2}=\frac{1}{2\left( 4\pi \right) ^{3}}\log \frac{3}{4}\text{ }\frac{1}{m}.
\end{equation}
The 4-loop and 5-loop integrals must be evaluated numerically. Four of
them can be reduced to integrals over a single momentum. The integrals $%
J_{3} $ and $J_{4e}$ already have this form. The integral $J_{4a}$ and $%
J_{4d}$ can be put into this form by first integrating over $p$:
\begin{eqnarray}
\int_{p}D_{1}^{2}\left( p\right) \left[ D_{1}\left( p+q\right) -D_{1}\left(
q\right) \right] &=&-\frac{3}{8\pi }\frac{m}{\left( q^{2}+m^{2}\right)
\left( q^{2}+4m^{2}\right) }, \\
\int_{p}D_{1}^{2}\left( p\right) \left[ D_{2}\left( p+q\right) -D_{2}\left(
q\right) \right] &=&\frac{1}{32\pi ^{2}}\frac{1}{mq}
\left[ \arctan \frac{q}{3m}-\arctan \frac{q}{2m}\right] .
\end{eqnarray}
Reducing the final momentum integrals to a 1-dimensional integral
and then evaluating it numerically, we obtain
\begin{eqnarray}
J_{3} &=&\frac{1}{\left( 4\pi \right) ^{5}}\left( -0.642143990\right) \frac{1%
}{m^{2}}, 
\\
J_{4a} &=&\frac{1}{\left( 4\pi \right) ^{6}}\left( -0.248926826\right) \frac{%
1}{m^{3}}, 
\\
J_{4d} &=&\frac{1}{\left( 4\pi \right) ^{6}}\left( 0.0165690298\right) \frac{%
1}{m^{3}}, 
\label{J4d}
\\
J_{4e} &=&\frac{1}{\left( 4\pi \right) ^{6}}\left( 0.00915691397\right) 
\frac{1}{m^{3}}.
\label{J4e}
\end{eqnarray}
The remaining two 5-loop integrals can be expressed as 3-loop integrals
with the Mercedes-Benz topology. The integral $I_{4b}$ in (\ref{A17})\
already has this form. The integral $I_{4c}$ in (\ref{A18})\ can be put in
this form by integrating over $p.$ These integrals can be reduced to
7-dimensional integrals that are then evaluated using the Monte Carlo method
to obtain
\begin{eqnarray}
J_{4b} &=&\frac{1}{\left( 4\pi \right) ^{6}}\left( -0.19\right) 
\frac{1}{m^{3}}, 
\\
J_{4c} &=&\frac{1}{\left( 4\pi \right) ^{6}}\left( -0.13\right) 
\frac{1}{m^{3}}.
\end{eqnarray}
If we use the prescription I of Ref.~\cite{Ramos2},
we must also evaluate the integral $J_0$ defined in (\ref{J0}):
\begin{equation}
J_{0}=-\frac{1}{4\pi }m.
\end{equation}

The integrals $I_{n}$ that survive in the large-$N$ limit can
all be reduced to integrals over a single momentum. After interchanging
orders of integration in (\ref{A22}) and integrating over $p$, the remaining
integral reduces to
\begin{equation}
I_{n}=-\frac{1}{8\pi ^{3}}\left( \frac{1}{8\pi m}\right)
^{n-1}\int_{0}^{\infty }dy\frac{y^{2}}{\left( y^{2}+1\right) ^{2}}
\left[ A(y) \right] ^{n-1}, 
\label{B24}
\end{equation}
where the function $A(y)$ is defined in (\ref{A-y}).
Applying the same strategy to (\ref{A23}), we obtain
\begin{equation}
J_{n}=-\frac{3}{16\pi ^{3}}\left( \frac{1}{8\pi m}\right)
^{n-1}\int_{0}^{\infty }dy\frac{y^{2}}{\left( y^{2}+1\right) \left(
y^{2}+4\right) }\left[ A(y) \right] ^{n-1}.
\label{B25}
\end{equation}

\newpage

\begin{center}
\begin{tabular}{r||c|c|c||c|c|c}
    & \multicolumn{3}{c||}{II}      & \multicolumn{3}{c}{III}  \\ 
\hline
    &             &    PMS     &          PMS        
    &             &    PMS     &          PMS           \\
$n$ & $\mu=1.039$ & $\mu$ real & $max |{\rm Im}\Delta|$
    & $\mu=0.947$ & $\mu$ real & $max |{\rm Im}\Delta|$ \\
\hline
 3  & 0.481 & 0.481 &                   & 0.456 & 0.456 &                   \\ 
 4  & 0.625 &       & $0.557\pm 0.156i$ & 0.601 &       & $0.525\pm 0.157i$ \\ 
 5  & 0.603 & 0.611 & $0.615\pm 0.245i$ & 0.574 & 0.579 & $0.574\pm 0.250i$ \\ 
 7  & 0.656 & 0.675 & $0.705\pm 0.339i$ & 0.627 & 0.641 & $0.653\pm 0.355i$ \\ 
11  & 0.709 & 0.742 & $0.818\pm 0.400i$ & 0.683 & 0.708 & $0.755\pm 0.438i$ \\ 
19  & 0.761 & 0.801 & $0.913\pm 0.416i$ & 0.737 & 0.769 & $0.851\pm 0.480i$ \\ 
35  & 0.809 & 0.850 & $0.971\pm 0.410i$ & 0.787 & 0.821 & $0.917\pm 0.493i$ \\ 
\end{tabular}
\end{center}
\begin{table}[tbp]
\caption{The values of $\Delta /(-Nu/96 \pi^2)$ in the large-$N$ limit 
at $n^{\rm th}$ order in the $\delta$ expansion 
using the prescriptions II and III
for various choices of $\mu$. 
Columns 1 and 4 are for fixed values of $\mu$.
Columns 2 and 5 are the values at the real solutions to the PMS criterion. 
Columns 3 and 6 are the values at the complex solutions 
with the maximal value of $|{\rm Im}\Delta|$.}
\label{tab1}
\end{table}

\newpage 

\begin{center}
\begin{tabular}{c|c|c|c}
$n$ & I  & II & III \\ 
\hline
2 & $\phantom{-}$1.314    &                       &                       \\ 
  & $-1.314$              &                       &                       \\ 
\hline 
3 &    $1.051\pm 0.710i$  &     0.192             &     0.182             \\ 
  & $-1.920$              &                       &                       \\ 
\hline 
4 & $\;0.679\pm 0.969i\;$ & $\;0.214\pm 0.084i\;$ & $\;0.200\pm 0.083i\;$ \\ 
  & $\phantom{-}$1.312    &                       &                       \\ 
  & $-2.270$              &                       &
\end{tabular}
\end{center}
\begin{table}[tbp]
\caption{The values of $\Delta /(- u/48 \pi^2) $ for $N=2$ 
using the prescriptions I, II, and III at $n^{\rm th}$ order in the LDE 
with $\mu$ evaluated at each of the solutions to the PMS criterion.}
\label{tab2}
\end{table}

\end{document}